%% file: conference_101719.tex
\documentclass[10pt,conference]{IEEEtran}
\usepackage{cite}
\usepackage{amsmath,amssymb,amsfonts}

\usepackage{textcomp}
\usepackage{xcolor}
\usepackage{multirow}
\usepackage{makecell}
\usepackage{subfigure}
\usepackage{graphicx}
\usepackage{array}
\newcolumntype{P}[1]{>{\centering\arraybackslash}p{#1}}
\usepackage{calc}
\usepackage{array}
\usepackage{optidef}
\usepackage{epstopdf}
\usepackage{mathtools}
\usepackage{mathrsfs}
\usepackage{bm}
\usepackage{cite}
\usepackage{boldline}
\usepackage{algorithm}
\usepackage{algorithmic}
\usepackage{amsmath}

\usepackage{booktabs}

\usepackage{siunitx}
\def\BibTeX{{\rm B\kern-.05em{\sc i\kern-.025em b}\kern-.08em
    T\kern-.1667em\lower.7ex\hbox{E}\kern-.125emX}}

\input{macros}

\newtheorem{remark}{Remark}

\begin{document}

\title{Channel Estimation for Reconfigurable Intelligent Surface Assisted Upper Mid-Band MIMO Systems}

\author{\IEEEauthorblockN{${\textrm{Jeongjae Lee}}$, ${\textrm{Chanwon Kim}}$, and ${\textrm{Songnam Hong}}$
}
\IEEEauthorblockA{
${\textrm{Department of Electronic Engineering, Hanyang University, Seoul, South Korea}}$
\\
Email: 
${\textrm{\{lyjcje7466, chanwon0721, snhong\}@hanyang.ac.kr}}$
}

\thanks{This work was supported in part by Institute of Information \& communications Technology Planning \& Evaluation (IITP) under the artificial intelligence semiconductor support program to nurture the best talents (IITP-2025-RS-2023-00253914) grant funded by the Korea government(MSIT) and in part by the National Research Foundation of Korea(NRF) grant funded by the Korea government(MSIT)(No. RS-2024-00409492).}}

\maketitle

\begin{abstract}
The upper mid-band (UMB) spectrum is a key enabler for 6G systems, yet reconfigurable intelligent surface (RIS)-assisted UMB communications face severe channel estimation challenges due to near-field propagation and transitional scattering, which induce strong spatial correlation and ill-conditioned least-squares (LS) formulations. To overcome this limitation, we propose a conditioning-aware channel estimation framework that transforms the inherently ill-conditioned high-dimensional problem into multiple well-conditioned subproblems via greedy column grouping. By systematically separating highly correlated RIS elements into distinct sub-blocks via piecewise RIS phase design, the proposed method directly improves Gram matrix conditioning and stabilizes piecewise LS reconstruction without relying on sparsity assumptions. Simulation results demonstrate that the proposed method significantly outperforms conventional LS and OMP-based estimators in pilot-limited and transitional UMB regimes, achieving robust performance with low computational complexity.

\end{abstract}

\begin{IEEEkeywords}
Reconfigurable intelligent surface (RIS), channel estimation, upper mid-band.
\end{IEEEkeywords}

\section{Introduction}

The upper mid-band (UMB) spectrum (7--24 GHz) has recently gained strategic importance as a key enabler for beyond-5G and emerging 6G systems \cite{Bazzi2026}. Positioned between sub-6 GHz and millimeter-wave (mmWave) bands, UMB offers a favorable trade-off between bandwidth and propagation robustness, making it attractive for high-capacity and wide-area deployments. Reconfigurable intelligent surfaces (RIS) further enhance UMB systems by dynamically controlling signal reflections to improve coverage and spatial degrees of freedom \cite{Kara2025}. In practical deployments such as fixed wireless access (FWA) and dense urban communications, RIS enables dynamic beam redirection to enhance coverage and signal-to-interference-plus-noise ratio (SINR). However, large-scale RIS operation at UMB frequencies fundamentally changes channel characteristics. Owing to the large RIS aperture relative to the wavelength, communication often occurs in the near-field region with spherical wavefront propagation, increasing channel estimation complexity.

Moreover, UMB channels typically operate in a transitional scattering regime with a moderate number of scatterers, and are therefore neither strictly sparse nor fully high-rank. In such environments, the angular sparsity assumed in compressed sensing (CS) becomes less pronounced, as near-field propagation spreads multipath energy across multiple angular bins, leading to grid mismatch and unreliable support detection \cite{chen2023channel,Yang2024}. Likewise, low-rank approximation methods degrade as the effective channel rank increases under moderate scattering \cite{chung2024efficient,Lee2025}. Two-timescale channel estimation (2TCE) improves pilot efficiency by exploiting the quasi-static RIS--BS link and applying least-square (LS)-based reconstruction for the user--RIS channel \cite{Hu2021,Lee2026}. However, its performance depends critically on Gram matrix conditioning. In RIS-assisted UMB systems, strong spatial correlation among RIS elements results in ill-conditioned Gram matrices within the reduced-pilot framework, making LS inversion highly sensitive to noise, especially in pilot-limited regimes.

To address these challenges, we propose a conditioning-aware channel estimation framework tailored for RIS-assisted UMB systems. Building upon the 2TCE architecture, the proposed method directly targets the fundamental conditioning bottleneck induced by a moderate number of scatters under near-field spatial correlation. The key contribution is a correlation-aware greedy column grouping strategy that transforms a high-dimensional ill-conditioned estimation problem into multiple well-conditioned subproblems. By redistributing highly correlated RIS elements across distinct sub-blocks via piecewise RIS phase design, the proposed approach improves Gram matrix conditioning and stabilizes piecewise LS reconstruction without relying on structural sparsity assumptions. Extensive simulations demonstrate that the proposed method achieves superior robustness over conventional LS and OMP-based estimators in transitional UMB regimes, while maintaining low computational complexity suitable for practical deployment.

{\em Notations.} Let $[N]\eqdef \{1,2,\dots,N\}$ for any integer $N$. We use $\av$ and $\Am$ to denote a column vector and matrix, respectively. Given a vector $\av\in\CC^{N}$, $\mbox{diag}(\av)$ denotes a diagonal matrix whose diagonal elements corresponds to the elements of $\av$. Given a $M \times N$ matrix $\Am$, let $\Am(i,:)$ and $\Am(:,j)$ denote the $i$-th row and $j$-th column of $\Am$, respectively. Given $m < M$ and $n < N$, let $\Am([m],:)$ and $\Am(:,[n])$ denote the submatrices by taking the first $m$ rows and $n$ columns of $\Am$, respectively. Also, $\Am^{\herm}$ and $\|\Am\|_2$ denote the Hermitian transpose and the $\ell_2$-norm, respectively. 
\section{Preliminaries}


We consider a RIS-assisted UMB (7--24 GHz) communication system, where a base station (BS) equipped with $N$ antennas serves single-antenna users within a cell. To enhance link reliability and coverage, a RIS with $M$ passive reflecting elements assists the communication between the user and the BS~\cite{Kara2025}. In this paper, we develop an efficient uplink channel estimation (CE) algorithm for RIS-assisted UMB systems. For clarity of exposition, we focus on a single-user scenario. The proposed method can be readily extended to multi-user settings by employing orthogonal pilot sequences as in~\cite{chen2023channel,Lee2025}. To focus on the cascaded user-RIS-BS channel, we assume that the direct user-BS channel is estimated using conventional approaches, e.g., by temporarily disabling the RIS \cite{chung2024efficient} or by suitable RIS phase manipulation \cite{chen2023channel}.

\subsection{Channel and Signal Model}
Let $\Fm\in\CC^{N\times M}$ denote the RIS-BS channel matrix and $\hv\in\CC^{M}$ denote the user-RIS channel vector. Note that due to a large antenna aperture for achieving a high spectral efficiency and a relatively large wavelength in UMB comparing with the mmWave or sub-THz bands, the near-field channel based on spherical wavefronts should be taken into account for a precise channel modeling \cite{Kara2025}. Therefore, we model the RIS-BS channel $\Fm$ and the user-RIS channel $\hv$ by following the near-field channel model in \cite{Yu2023}, where $L^{\rm RB}$ and $L^{\rm UR}$ denote the number of scatters in the RIS-BS channel and the user-RS channel, respectively. The CE procedure spans $T$ (pilot) time slots. At time slot $t \in [T]$, the user transmits a pilot symbol $s_t\in\CC$ satisfying $\EE\left[|s_t|^2\right]=1,\forall t$.  Then, the received signal at time slot $t$ is given by
\begin{align}
    \yv_t &= \sqrt{P}\Fm\diag\left(\psiv_t\right)\hv s_{t} + \nv_t, \label{eq:received}
\end{align} where $P$ is the transmit power, $\psiv_t\in\CC^{M}$ is the RIS reflection vector satisfying the constant-modulus constraint $|\psiv_t(m)|=1,\forall m\in[M]$, and $\nv_t\sim\Cc\Nc(\mathbf{0},\sigma^2\Id_N)$ denotes additive white Gaussian noise with variance $\sigma^2$.

\subsection{Two-Timescale Channel Estimation Framework}

In UMB systems, both $\Fm$ and $\hv$ typically exhibit {\em moderate-to-rich} scattering characteristics, i.e., $L^{\rm RB}\gg 1$ and $L^{\rm UR}\gg 1$, placing the channel in a transitional regime between highly sparse mmWave bands and rich-scattering sub-6 GHz bands~\cite{Kara2025}. In such environments, strict sparsity or rank-deficiency assumptions become unreliable, limiting the effectiveness of 
conventional compressed sensing (CS) \cite{chen2023channel,Yang2024} and low-rank approximation methods \cite{chung2024efficient,Lee2025}. To cope with this structural ambiguity while maintaining manageable pilot overhead, we adopt the two-timescale channel estimation (2TCE) framework, introduced in \cite{Hu2021,Lee2026}. The key insight of 2TCE is to exploit the disparity in channel coherence times between the RIS-BS channel $\Fm$ and the user-RIS channel $\hv$.
Specifically, the RIS-BS channel remains quasi-static over a long coherence interval, whereas the user-RIS channel varies more rapidly due to user mobility. Accordingly, $\Fm$ is first estimated using sufficient pilot resources via dual-link training and then reused to reduce the pilot overhead required for estimating $\hv$ \cite{Hu2021}.


Let $\hat{\mathbf{F}}$ denote the estimated RIS-BS channel and define $\hat{\mathbf{F}}_t = \hat{\mathbf{F}}\,\mathrm{diag}(\boldsymbol{\psi}_t)$,
where $\boldsymbol{\psi}_t$ is randomly generated across time slots. The user-RIS channel is then estimated via least-squares (LS):
\begin{equation}
    \hat{\hv} = \left(\sum_{t=1}^{T}\hat{\Fm}_t^{\herm}\hat{\Fm}_t\right)^{-1}\sum_{t=1}^{T}\hat{\Fm}_t^{\herm}\tilde{\yv}_t,\label{eq:LS}
\end{equation} where $\tilde{\yv}_t = {{\yv}_t}/{\sqrt{P}s_t}\in\CC^{N}$. In \cite{Hu2021}, the minimum pilot overhead required to ensure full column rank is given by $T=\lceil\frac{M}{N}\rceil$, which guarantees that the Gram matrix $\sum_{t=1}^{T}\hat{\mathbf{F}}_t^{\herm}\hat{\mathbf{F}}_t$ is invertible. For example, when $N = 64$ and $M = 256$, the required pilot length reduces to $T = 4$, enabling the estimation of the $256$-dimensional vector $\mathbf{h}$ with only four pilot symbols.

Despite its pilot efficiency, the LS-based 2TCE framework has two notable limitations. 
First, its performance relies on favorable conditioning of the Gram matrix. In practical RIS-assisted UMB systems, strong spatial correlation among RIS elements—especially under near-field propagation—can render the Gram matrix severely ill-conditioned, leading to significant noise amplification in the LS inversion. 
Second, when the RIS dimension becomes large (e.g., $M$ on the order of several hundreds), the required matrix inversion incurs $\mathcal{O}(M^3)$ computational complexity, imposing a substantial implementation burden for large-scale RIS deployments. These limitations reveal a fundamental gap in existing 2TCE-based methods: while pilot overhead can be reduced via timescale separation, numerical stability is not explicitly addressed. Therefore, an enhanced channel estimation framework is required that preserves pilot efficiency while explicitly improving the conditioning of the sensing structure in RIS-assisted UMB environments.

%

\section{Problem Formulation}
To overcome the conditioning limitations of the conventional LS-based 2TCE framework in UMB systems, we adopt the piecewise RIS phase design proposed in \cite{Lee2026}. Our main idea is to partition the RIS elements into multiple disjoint groups such that the high-dimensional estimation problem can be decomposed into structured lower-dimensional sub-problems. Specifically, 
the RIS is divided into $Q$ disjoint groups. Let $\Mc_q \subseteq [M]$ denotes the index set of RIS elements in group $q\in[Q]$, satisfying
\begin{equation}
\Mc_q \cap \Mc_{q'} = \emptyset,\quad
\bigcup_{q=1}^{Q} \Mc_q = [M].\label{eq:const1}
\end{equation}
Without loss of generality, we assume equal group sizes, 
\begin{equation}
|\Mc_q| = \frac{M}{Q} = M', \quad \forall q,\label{eq:const2}
\end{equation}
where $M'$ is an integer. Under this partitioning, the received signal in \eqref{eq:received} can be equivalently expressed as
\begin{equation}
    \yv_t = \sqrt{P}\sum_{q=1}^{Q}\Fm_q\mbox{diag}\!\left(\psiv_{t,q}\right)\!\hv_q s_{t} + \nv_{t},
\end{equation} where $\Fm_q=\Fm(:,\Mc_q)\in\CC^{N\times M'}$, $\psiv_{t,q}=\psiv_{t}(\Mc_q)\in\CC^{M'}$, and $\hv_q=\hv(\Mc_q)\in\CC^{M'}$. This reformulation decomposes the original $M$-dimensional channel estimation problem into $Q$ coupled sub-problems of dimension $M'$, whose numerical stability critically depends on the grouping strategy $\{\Mc_q\}_{q=1}^Q$.

\begin{figure}[t]
\centering
\includegraphics[width=0.95\linewidth]{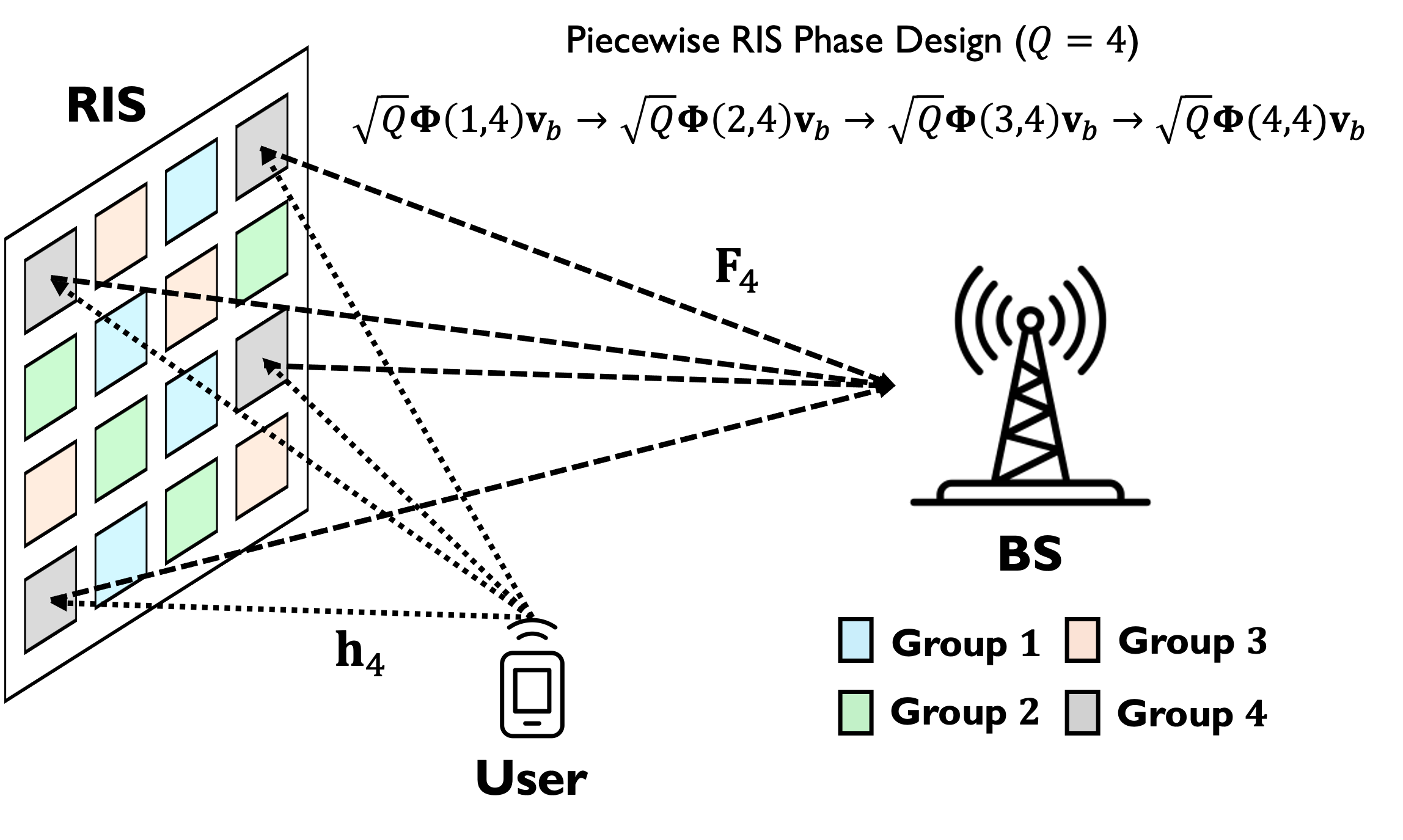}
\caption{Illustration of the RIS-assisted UMB system and piecewise RIS phase design, e.g., $M=16$ and $Q=4$.} 
\label{fig:pwRIS}
\end{figure}

\subsection{Piecewise RIS Phase Design}

The objective of the piecewise RIS phase design is to generate $Q$ decoupled observations, as illustrated in Fig.~\ref{fig:pwRIS}, each corresponding to a single RIS group, such that the high-dimensional channel estimation problem can be decomposed into independent lower-dimensional sub-problems. To this end, the total $T$ pilot slots are divided into $B\leq M'$ subframes, each containing $Q$ time slots, i.e., $T = QB$. Within each subframe $b \in [B]$, orthogonal phase patterns are applied across the $Q$ groups to enable inter-group decoupling. Specifically, for time slot $k\in[Q]$ within subframe $b$, the RIS reflection vector associated with group $q$ is designed as
\begin{equation}
    \psiv_{k+(b-1)Q,q} = \sqrt{Q}\Phim(k,q)\mathbf{v}_b,
\end{equation} where $\Phim \in \CC^{Q\times Q}$ is a unitary Hadamard matrix satisfying $\Phim\Phim^{\herm}=\mathbf{I}$, and $\mathbf{v}_b\in\CC^{M'}$ denotes the $b$-th column of an $M'\times M'$ Hadamard matrix. The orthogonality of $\Phim$ ensures that, after linear processing across the $Q$ time slots within each subframe, the contributions from different RIS groups become separable. After collecting the $Q$ pilot observations within subframe $b$, we stack them as
\begin{align}
    \Ym_b &= \begin{bmatrix}
        \tilde{\yv}_{1+(b-1)Q} &\cdots &  \tilde{\yv}_{bQ}
    \end{bmatrix}\\
    &=\sqrt{Q}\begin{bmatrix}
        \Fm_1\mbox{diag}\!\left(\mathbf{v}_b\right)\!\hv_1 &\cdots &\Fm_Q\mbox{diag}\!\left(\mathbf{v}_b\right)\!\hv_Q
    \end{bmatrix}\Phim + {\Nm}_b,\nonumber
\end{align} where
${\Nm}_b = \left[\frac{\nv_{1+(b-1)Q}}{\sqrt{P}s_{1+(b-1)Q}}\;\cdots\;\frac{\nv_{bQ}}{\sqrt{P}s_{bQ}}\right]\in\CC^{N\times Q}$. 
To decouple the contributions from different RIS groups, we right-multiply and normalize $\Ym_b$ by the unitary matrix $\Phim^{\herm}$ and by $Q$, yielding $\tilde{\mathbf{Y}}_b
=
\frac{\mathbf{Y}_b \Phim^{\herm}}{\sqrt{Q}}
\in \mathbb{C}^{N \times Q}$ and $\tilde{\mathbf{N}}_b
=
\frac{\mathbf{N}_b \Phim^{\herm}}{\sqrt{Q}}
\in \mathbb{C}^{N \times Q}$.
Using the property $\Phim \Phim^{\herm} = \mathbf{I}$, we obtain
\begin{equation}
    \mathbf{z}_{b,q}
    =
    \tilde{\mathbf{Y}}_b(:,q)
    =
    \mathbf{F}_q
    \mathrm{diag}(\mathbf{v}_b)
    \mathbf{h}_q
    +
    \tilde{\mathbf{n}}_{b,q},
\end{equation}
where $\tilde{\mathbf{n}}_{b,q} = \tilde{\mathbf{N}}_b(:,q) \in \mathbb{C}^{N}$.
Since $\mathbf{\Phi}$ is unitary, the noise remains circularly symmetric complex Gaussian with covariance scaled by $1/Q$.



\subsection{RIS Grouping Optimization Problem}
By stacking the $B$ decoupled observations $\{\mathbf{z}_{b,q}\}_{b=1}^{B}$ corresponding to group $q$, the LS estimator of $\mathbf{h}_q$ is obtained as
\begin{equation}
    \hat{\mathbf{h}}_q
    =
    \left(
        \sum_{b=1}^{B}
        \hat{\mathbf{F}}_{b,q}^{\herm}
        \hat{\mathbf{F}}_{b,q}
    \right)^{-1}
    \sum_{b=1}^{B}
        \hat{\mathbf{F}}_{b,q}^{\herm}
        \mathbf{z}_{b,q}, \label{eq:group_ls}
\end{equation}
where $\hat{\Fm}_{b,q} = \hat{\Fm}(:,\Mc_q)\mbox{diag}\!\left(\mathbf{v}_b\right)$. 
The estimation accuracy of \eqref{eq:group_ls} critically depends on the conditioning of the group-wise Gram matrix
\begin{equation}
\mathbf{G}_q
=
\sum_{b=1}^{B}
\hat{\mathbf{F}}_{b,q}^{\herm}
\hat{\mathbf{F}}_{b,q}.\label{eq:gwGram}
\end{equation} To improve the conditioning of the group-wise Gram matrices, we design a correlation-aware grouping strategy for the RIS elements, denoted by $\Mc_q,\forall q$. Under the piecewise RIS phase design, each group $q$ yields a reduced-dimensional LS subproblem of size $M'=M/Q$, whose associated Gram matrix $\Gm_q$ is defined in \eqref{eq:gwGram}. The numerical stability of the LS estimator is critically determined by the condition number of the Gram matrix $\Gm_q$, denoted as $\eta(\mathbf{G}_q)$. In RIS-assisted UMB environments, adjacent RIS elements typically exhibit strong spatial correlation due to a high but finite number of scatters. If highly correlated columns are grouped together, $\Gm_q$ becomes poorly conditioned, resulting in significant noise amplification during matrix inversion. Therefore, an effective grouping strategy should distribute weakly correlated (spatially separated) RIS elements within the same group to reduce intra-group correlation and enhance the conditioning of $\Gm_q$.

Importantly, the grouping strategy only needs to be optimized once based on the estimated quasi-static RIS-BS channel $\hat{\Fm}$ and can be reused for subsequent user-RIS channel estimations within its coherence time. Moreover, since each subproblem has dimension $M' = M/Q \ll M$ when $Q \ge 2$, the LS inversion is performed on $Q$ independent matrices of size $M' \times M'$ instead of a single $M \times M$ matrix. As a result, the computational complexity is reduced from $\mathcal{O}(M^3)$ to $\mathcal{O}(QM'^3)=\mathcal{O}\left({M^3}/{Q^2}\right)$, which significantly improves scalability for large-scale RIS deployments.
Based on the above observations on Gram matrix conditioning, we formulate the RIS grouping problem as 
\begin{align*}
    \mathcal{P}_0:\quad
    \min_{\{\mathcal{M}_q\}_{q=1}^{Q}}
    &\;
    \max_{q\in[Q]}
    \eta(\mathbf{G}_q)
    \\
    \text{s.t.}\quad & \text{\eqref{eq:const1} and \eqref{eq:const2}}.
\end{align*} Problem $\mathcal{P}_0$ aims to determine a partition of the RIS elements that minimizes the worst-case condition number across all groups, thereby ensuring uniformly stable LS subproblems. 




\section{Proposed Greedy Column Grouping Algorithm}

The optimization problem $\mathcal{P}_0$ is combinatorial in nature, as it seeks an equal-cardinality partition of $M$ RIS elements into $Q$ disjoint groups. The number of feasible partitions grows exponentially with $M$, rendering exhaustive search computationally prohibitive for large-scale RIS systems. Therefore, obtaining the globally optimal solution is impractical in realistic UMB deployments. To address this challenge, we develop a low-complexity greedy column grouping algorithm for the estimated RIS-BS channel $\hat{\Fm}$ that efficiently approximates the solution of $\mathcal{P}_0$.

\subsection{Surrogate Formulation and Rationale}




To obtain a tractable and robust alternative of the problem $\Pc_0$, we exploit the fundamental relationship between Gram matrix conditioning and column correlation. Specifically, the Gram matrix $\mathbf{G}_q$ can be expressed as the inner-product matrix of the columns of $\hat{\mathbf{F}}_{b,q}$, whose off-diagonal entries are determined by pairwise correlations. High mutual correlation among columns implies spatial redundancy, reduces the effective rank of $\mathbf{G}_q$, and degrades its conditioning, thereby amplifying noise in LS inversion. Motivated by this observation, we introduce the following surrogate optimization problem:
\begin{align}\label{eq:surrogate}
    \mathcal{P}_1:\quad\min_{\{\mathcal{M}_q\}_{q=1}^{Q}}
    &\max_{q\in[Q]}
    \sum_{i<j \in \mathcal{M}_q}
    w_{i,j},\nonumber\\
    \text{s.t.}\quad & \text{\eqref{eq:const1} and \eqref{eq:const2}},
\end{align}
where the normalized mutual correlation between the $i$-th and $j$-th columns of $\mathbf{F}$ is defined as
\begin{equation}
    w_{i,j}
    =
    \frac{\left|\mathbf{F}(:,i)^{\herm}\mathbf{F}(:,j)\right|}
    {\left\|\mathbf{F}(:,i)\right\|_2 \left\|\mathbf{F}(:,j)\right\|_2}.\label{eq:cor}
\end{equation}



\subsection{Greedy Partitioning Strategy and Implemenation}
Despite the introduction of the surrogate problem $\Pc_1$, the combinatorial nature of $\Pc_1$ is still challenging. To address this, we adopt a greedy strategy that sequentially constructs the groups while explicitly separating highly correlated columns.

\subsubsection{Seed Initialization}

Without loss of generality, we assume that the number of groups $Q$ is even. The objective of this phase is to identify the $Q$ columns that exhibit the strongest pairwise correlations and ensure that they are assigned to different groups.
\begin{itemize}
    \item \textbf{Correlation Analysis:} Compute the mutual correlation $w_{i,j}$ for all $i \neq j$, and sort the pairs in descending order.
    \item \textbf{Seed Separation:} Select the $Q/2$ largest non-overlapping pairs to guarantee that no column index appears in more than one selected pair.
    \item \textbf{Group Initialization:} The resulting $Q$ distinct indices are used to initialize the groups $\{\mathcal{M}_1,\dots,\mathcal{M}_Q\}$. 
\end{itemize}

\subsubsection{Sequential Greedy Assignment}
After the seed initialization, the remaining $M - Q$ columns are assigned sequentially. At each step, the goal is to place an assigned column into the group where it incurs the smallest additional intra-group correlation.
\begin{itemize}
    \item \textbf{Cumulative Correlation Evaluation:} For each unassigned column $c' \notin \Omega$, compute its average cumulative correlation with the existing members of each group $\mathcal{M}_q$:
        \begin{equation}
        \frac{1}{|\mathcal{M}_q|}
        \sum_{c \in \mathcal{M}_q}
        w_{c',c}.\label{eq:C}
    \end{equation} This quantity measures the incremental correlation introduced if column $c'$ were assigned to group $q$.

    \item \textbf{Greedy Assignment:} Column $c'$ is assigned to the group that minimizes the incremental correlation:
    \begin{equation}\label{eq:qstar}
        q^\star
        =
        \arg\min_{q \in [Q]} \frac{1}{|\mathcal{M}_q|}
        \sum_{c \in \mathcal{M}_q}
        w_{c',c},
    \end{equation} followed by the update $\mathcal{M}_{q^\star} \leftarrow \mathcal{M}_{q^\star} \cup \{c'\}$.

    \item \textbf{Iteration:} The index set $\Omega$ is updated accordingly, and the procedure repeats until all $M$ RIS elements are assigned.
\end{itemize}


\subsubsection{Implementation}
Letting $\Mc_q^{\rm opt}$ denote the optimized group $q$ via the proposed greedy colum grouping algorithm, the optimized group-wise Gram matrix in \eqref{eq:gwGram} is expressed as $\Gm_q^{\rm opt}$. Then, the LS estimate of $\hv_q$ is obtained as
\begin{equation}
    \hat{\mathbf{h}}_q^{\rm opt}
    =
    \left(
        \Gm_q^{\rm opt}
    \right)^{-1}
    \sum_{b=1}^{B}
        \left(\hat{\Fm}_{b,q}^{\rm opt}\right)^{\herm}
        \mathbf{z}_{b,q},
\end{equation} where $\hat{\Fm}_{b,q}^{\rm opt} = \hat{\mathbf{F}}(:,\mathcal{M}_q^{\rm opt})\mbox{diag}(\mathbf{v}_b)$.

{

\begin{remark}[Computational Complexity Analysis]
The computational complexity of the proposed greedy column grouping is dominated by the pairwise correlation computation in \eqref{eq:cor}, resulting in $\mathcal{O}(NM^2)$ operations. Since the grouping depends only on the quasi-static RIS--BS channel $\hat{\mathbf{F}}$, it needs to be performed only once within its coherence interval. Therefore, this offline complexity is not included in the runtime complexity of the user--RIS channel estimation. Compared with conventional LS-based 2TCE, which requires a matrix inversion with complexity $\mathcal{O}(M^3)$, the proposed piecewise LS reduces the inversion complexity to $\mathcal{O}(M^3/Q^2)$. In contrast, OMP-based estimators typically incur higher iterative complexity due to repeated support selection and residual updates, making the proposed method computationally more efficient in practical RIS-assisted UMB systems.
\end{remark}
}

\begin{figure*}[t]
\centering
\includegraphics[width=1\linewidth]{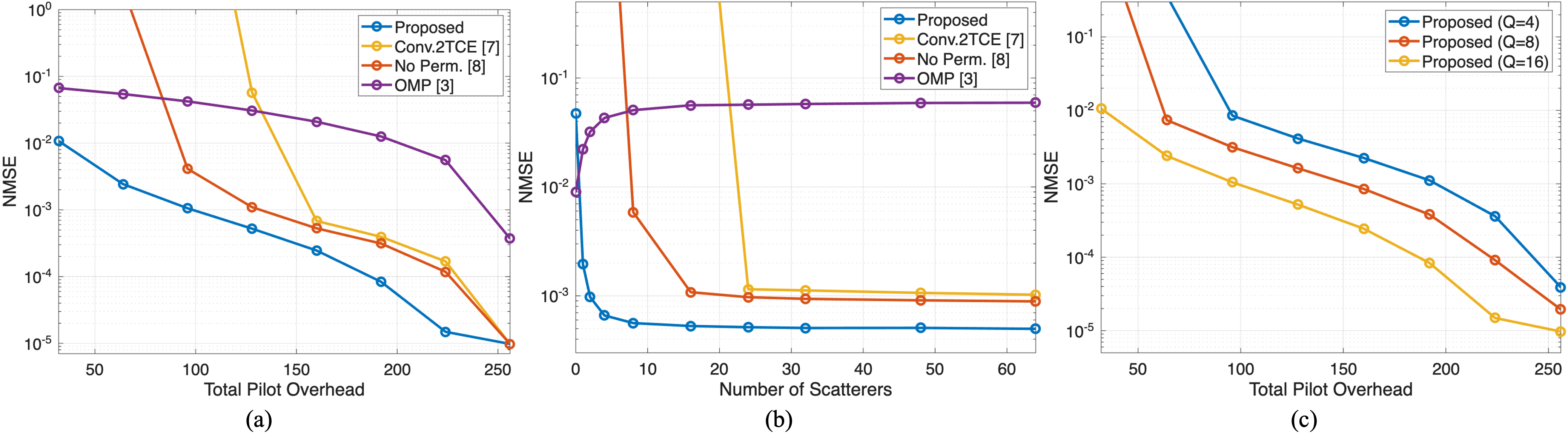}
\caption{NMSE performance of different channel estimation methods in the RIS-assisted UMB system. 
(a) NMSE versus the total pilot overhead. $L^{\rm RB}=L^{\rm UR}=16$. 
(b) NMSE versus the number of scatterers with a fixed pilot overhead. $Q=16$ and $B=8$.
(c) NMSE versus the total pilot overhead for different numbers of RIS groups $Q$. $L^{\rm RB}=L^{\rm UR}=16$.} 
\label{fig:NMSE_results}
\end{figure*}




\section{Simulation Results}

In this section, we evaluate the performance of the proposed channel estimation method through extensive numerical simulations. We consider an RIS-assisted UMB system operating at a carrier frequency of $f_c = 15$ GHz, representative of the upper mid-band (7--24 GHz) spectrum. The BS is located at the origin $(0, 0, 0)$, while the RIS is deployed at $(0, 20, 10)$ m. The user is randomly distributed within a circular region centered at $(40, 20, 0)$ m with a radius of $15$ m. The system employs $N=64$ and $M=256$. The signal-to-noise ratio (SNR) is set to $20$ dB throughout the simulations, and the normalized mean squared error (NMSE), averaged over 2000 independent channel realizations, is used as the performance metric.

Fig.~\ref{fig:NMSE_results} illustrates the NMSE performance under the considered RIS-assisted UMB system. As shown in Fig.~\ref{fig:NMSE_results}(a), the proposed greedy column grouping consistently achieves the lowest NMSE across all pilot regimes, achieving about $10^{-2}$ accuracy with only $32$ pilot signals. In contrast, the conventional 2TCE (Conv.2TCE \cite{Hu2021}) suffers from severe degradation in the low-pilot regime due to the ill-conditioned Gram matrix caused by strong spatial correlation among RIS elements. Although the piecewise LS without column grouping (No Perm. \cite{Lee2026}) reduces the dimensionality of each LS subproblem, spatially adjacent RIS elements remain grouped together, resulting in highly correlated sensing matrices and degraded estimation accuracy. Furthermore, the OMP-based estimator (OMP \cite{chen2023channel}) shows limited improvement since the considered UMB channel does not strictly satisfy the angular sparsity assumption. Fig.~\ref{fig:NMSE_results}(b) further evaluates the robustness with respect to the number of scatterers, where the conventional LS-based estimators significantly degrade in sparse scattering environments due to near-field spatial correlation, whereas the proposed greedy grouping maintains stable NMSE by distributing highly correlated RIS elements across different groups. Finally, Fig.~\ref{fig:NMSE_results}(c) shows that increasing the number of RIS groups $Q$ improves the NMSE performance since each subproblem becomes smaller ($M'=M/Q$), leading to better-conditioned Gram matrices. These results confirm that the proposed correlation-aware grouping effectively improves the numerical stability of LS reconstruction and provides robust channel estimation performance in practical RIS-assisted UMB systems.

\section{Conclusion}
We proposed a conditioning-aware channel estimation framework for RIS-aided communication systems operating in the upper mid-band (UMB). By introducing a correlation-aware greedy column grouping strategy, highly correlated RIS elements are distributed across different sub-blocks, improving Gram matrix conditioning and stabilizing piecewise LS reconstruction. 
Simulation results demonstrated significant NMSE gains in pilot-limited and sparsely scattered regimes. Compared with conventional 2TCE and OMP-based estimators, the proposed method exhibits superior robustness in transitional UMB environments where strict sparsity assumptions are no longer valid. Beyond channel estimation, the proposed correlation-aware grouping principle offers a general structural design framework for large-scale near-field systems, including multi-user interference management and near-field beamforming, where conditioning and spatial de-correlation are critical.



\bibliographystyle{IEEEtran}
\bibliography{CS_REF}

\end{document}

%% file: macros.tex
\long\def\comment#1{}

\newfont{\bbb}{msbm10 scaled 700}

\newfont{\bb}{msbm10 scaled 1100}
\newcommand{\CC}{\mbox{\bb C}}

\newcommand{\EE}{\mbox{\bb E}}

\newcommand{\av}{{\bf a}}

\newcommand{\hv}{{\bf h}}

\newcommand{\nv}{{\bf n}}

\newcommand{\yv}{{\bf y}}

\newcommand{\Am}{{\bf A}}

\newcommand{\Fm}{{\bf F}}
\newcommand{\Gm}{{\bf G}}

\newcommand{\Id}{{\bf I}}

\newcommand{\Nm}{{\bf N}}

\newcommand{\Ym}{{\bf Y}}

\newcommand{\Cc}{{\cal C}}

\newcommand{\Mc}{{\cal M}}
\newcommand{\Nc}{{\cal N}}

\newcommand{\Pc}{{\cal P}}

\newcommand{\psiv}{\hbox{\boldmath$\psi$}}

\newcommand{\Phim}{\hbox{\boldmath$\Phi$}}

\newcommand{\diag}{{\hbox{diag}}}

\renewcommand{\arg}{{\hbox{arg}}}

\newcommand{\eqdef}{\stackrel{\Delta}{=}}

\newcommand{\herm}{{\sf H}}
